\documentclass[a4paper,11pt]{article}
\pdfoutput=1 

\usepackage{jcappub} 
\usepackage[utf8]{inputenc}
\usepackage{amsmath}
\usepackage{amsfonts}
\usepackage{amssymb}
\usepackage{mathtools}
\usepackage{bm}
\usepackage{xcolor}
\usepackage{soul}
\usepackage[numbers,sort&compress]{natbib}

\newcommand{\be}{\begin{equation}}
\newcommand{\ee}{\end{equation}}
\newcommand{\bea}{\begin{eqnarray}}
\newcommand{\eea}{\end{eqnarray}}

\DeclareMathAlphabet{\mathcal}{OMS}{cmsy}{m}{n}

\title{Revisiting $f(R)$ gravity's rainbow: Inflation and primordial fluctuations}

\author[a]{Yoelsy Leyva,}
\author[a]{Giovanni Otalora}

\affiliation[a]{Departamento de F\'isica, Facultad de Ciencias, Universidad de Tarapac\'a, Casilla 7-D, Arica, Chile \label{addr1}}

\emailAdd{yoelsy.leyva@academicos.uta.cl }
\emailAdd{giovanni.otalora@academicos.uta.cl}

\abstract{We study inflation and the generation of primordial fluctuations in $f(R)$ gravity's rainbow. We calculate the cosmological perturbations and then the scalar and tensor primordial power spectrum. We contrast the predictions of the model with the current observational data from PLANCK and BICEP/Keck. Particularly, we found new results for the scalar spectral index $n_s$ and the tensor-to-scalar ratio $r$ along with new observational constraints on the rainbow functions. 
 }

\begin{document}
\maketitle
\flushbottom


\section{Introduction}\label{Introduction}
A quasi-exponential accelerating phase, before the radiation decelerating era, that corresponds to cosmic inflation is widely accepted as the standard paradigm of the early Universe.  It can give a solution to the several long-standing puzzles of the Hot Big-Bang standard cosmological model \cite{Starobinsky,Guth:1980zm,Albrecht:1982wi,Linde:1981mu}. Furthermore, the most intriguing feature of inflation, is that it gives us a causal interpretation of the origin of the Cosmic Microwave Background (CMB) temperature anisotropies, and it provides us with a mechanism to explain the Large-Scale Structure (LSS) \cite{Weinberg:2008zzc}. This mechanism is based on the generation of quantum fluctuations which are amplified in physical scale during inflation, leading to a Gaussian, scale-invariant and adiabatic primordial density perturbations \cite{Weinberg:2008zzc,Baumann:2018muz}.  The measure of the scale dependence of the power spectrum of curvature perturbation is given by the scalar spectral index $n_{s}$, which is tightly constrained by the latest PLANCK data \cite{Planck:2018jri} to be $n_{s}=0.9649\pm 0.0042$ (at $68\%$  C.L). This result indicates that the primordial power spectrum is nearly scale independent. A further prediction of inflationary models is the generation of tensor perturbations as a background of primordial gravitational waves (PGWs), whose amplitude can be parameterized in terms of the tensor-to-scalar ratio $r$ \cite{Maggiore:2018sht}. As for the tensor-to-scalar ratio, we have not detected tensor perturbations until now. New data from BICEP/Keck 2021 \cite{BICEPKeck:2021gln} have been published, leading to a considerably stronger upper bound on r: $r_{0.05}=0.014^{+0.010}_{-0.011}$ ($r_{0.05}<0.036$ (at $95\%$ C.L.)), in comparison to PLANCK $2018$ data \cite{Planck:2018jri}.

Several different models have been studied to explain the physics of inflation. Among them we have $f(R)$ gravity that includes Starobinsky inflation \cite{Starobinsky}, as well as generalized scalar-tensor models \cite{Faraoni:2004pi} that encompass minimally and non-minimally coupled scalar fields models. A non-minimal coupling arises from a variety of model-building efforts, e.g., in supergravity and string theory, and it is also required as a counterterm when considering renormalization of scalar fields in curved spacetimes \cite{Birrell:1982ix}. Both $f(R)$ gravity and scalar fields models have been widely studied in the literature and their predictions for inflation are very well known. Starobinsky inflation is currently the best inflationary model in explaining the latest observations \cite{Planck:2018jri,BICEPKeck:2021gln}. On the other hand, scalar fields models have traditionally been used as the standard prototype for inflationary models. Although, the simplest and well-known examples of scalar potentials (the monomial power-law inflaton potentials: quadratic chaotic potential and quartic potential), have been strongly disfavored by current observations \cite{BICEPKeck:2021gln}, scalar field inflationary models may still be compatible with observations for other kind of potentials (asymptotically flat plateau like potentials). This is the case of the Einstein frame potentials associated with the Starobinsky model \cite{Starobinsky}, non-minimally coupled Higgs inflation \cite{Fakir:1990eg,Bezrukov:2007ep,Mishra:2018dtg,Mishra:2022ijb}, T-model and E-model $\alpha$-attractors \cite{Kallosh:2013hoa,Kallosh:2013yoa,Kallosh:2021mnu}, and the D-brane KKLT potential \cite{Kallosh:2021mnu,Kachru:2003aw,Kallosh:2019hzo,Kachru:2003sx}.  Another alternative class of modified gravity theories that have recently been addressed in the literature in explaining inflation are the torsional modified gravity theories \cite{Cai:2015emx}. These theories are extensions of the so-called teleparallel gravity \cite{Einstein,TranslationEinstein,Early-papers1,Early-papers2,Early-papers3,Early-papers4,Early-papers5,Early-papers6,Aldrovandi-Pereira-book,AndradeGuillenPereira-00,Arcos:2005ec}. The most popular examples are $f(T)$ gravity \cite{Ferraro:2006jd,Linder:2010py,Gonzalez-Espinoza:2018gyl,Harko:2014sja,Harko:2014aja}, scalar-torsion gravity \cite{Geng:2011aj,Geng:2011ka,Xu:2012jf,Otalora:2013tba,Otalora:2013dsa,Otalora:2014aoa,Skugoreva:2014ena,Gonzalez-Espinoza:2021mwr,Gonzalez-Espinoza:2020jss}, and higher-order teleparallel gravity \cite{Otalora:2016dxe,Gonzalez-Espinoza:2021nqd}. The calculation of the scalar and tensor power spectrum for generalized scalar-torsion theories taking into account the breaking of local Lorentz symmetry found in these theories has been performed in Ref. \cite{Gonzalez-Espinoza:2020azh}. A reconstruction scheme for inflation through the parametrization (or attractor) of the inflationary observables $n_s$ and $r$ as functions of the number of $e$-folds $N$ has been carried out in Ref \cite{Gonzalez-Espinoza:2021qnv}.

Gravity's Rainbow is an interesting proposal for a small-scale, Ultra-Violet (UV) modification of GR, and keeping GR as the low-energy limit \cite{Magueijo:2002xx}. Just as in other quantum gravity theories, the origin of Gravity's Rainbow is related to the non-renormalizability of GR and the difficulties that arise when trying to quantize gravity \cite{Stelle:1976gc}. Nevertheless, unlike what occurs in other quantum gravity frameworks \cite{Horava:2009uw}, in Gravity's Rainbow the ultraviolet (UV) modification appears directly in the spacetime metric. Since it is an extension of double special relativity to curved spacetimes the usual energy-momentum relations are modified through new contributing terms that depend on the probe energy. Thus, in Gravity's Rainbow the metric tensor is modified near the Planck scale which becomes a function of energy of the particle probing the spacetime \cite{Magueijo:2002xx}. Furthermore, the study of inflation in the context of $f(R)$ gravity with an energy-dependent spacetime metric, the so-called $f(R)$ gravity's rainbow, has already been performed in the literature. In Ref. \cite{Chatrabhuti:2015mws} (see also Ref. \cite{Channuie:2019kus}) the authors concentrated on Starobinsky model with the rainbow functions being power-law functions of the Hubble rate and they compared their results for $n_s$ and $r$ with observational data. Later in Ref. \cite{Waeming:2020rir} the analysis was extended to the logarithmic-corrected $R^2$ model and the Einstein-Hu-Sawicki model and also performing a comparison with the latest Planck 2018 data. Finally, the study of slow-roll inflation in the framework of $f(T)$ gravity's rainbow was achieved in Ref. \cite{Leyva:2021fuo}. In this latter article the authors carried out a full calculation of the inflationary primordial perturbations and they contrasted their results with the latest data from Planck $2018$ and BICEP/Keck $2021$.

In the present paper we study inflation and the generation of primordial fluctuations in $f(R)$ gravity's rainbow. Our goal is to carry out a detailed investigation of the slow-roll dynamics and the modifications to the scalar and tensor power spectrum sourced by the rainbow functions that arise in the effective spacetime metric. The manuscript is organized as follows: in Section \ref{Sec1}, we review the background equations for $f(R)$ gravity's rainbow. In Section \ref{Sec2} we establish the general setup for slow-roll inflation. In Section \ref{Sec3}  we study the evolution of the primordial scalar and tensor fluctuations by perturbing the modified field equations and by obtaining the Mukhanov-Sasaki equation. In Section \ref{Sec4} we apply our general results to a concrete inflationary model and confront its predictions with the current PLANCK and BICEP/Keck data. Finally, in Section \ref{conclusion_f} we give the conclusions. 
 
\section{$f(R)$ Gravity's Rainbow}\label{Sec1}

One starts from the action
\be
S=\frac{1}{2 \kappa^2}\int{d^{4}{x} \sqrt{-g} f(R)}+S_{m},
\label{action1}
\ee where $\kappa^2=8 \pi G$, $f(R)$ is a function of the Ricci scalar and $S_{m}$ is action of matter.

Varying this action with respect to the spacetime metric one obtains the following field equations
\be
F(R) R_{\mu \nu}-\frac{1}{2} f(R) g_{\mu \nu}-\nabla_{\mu}\nabla_{\nu}{F(R)}+g_{\mu \nu}\Box{F(R)}=\kappa^2 \mathcal{T}_{\mu \nu},
\label{FEQ}
\ee where $F(R)\equiv \partial{f}/\partial{R}$, $\Box=g^{\mu\nu}\nabla_{\mu}\nabla_{\nu}$ is Laplacian-Beltrami operator, and 
\be
\mathcal{T}_{\mu \nu}=-\frac{2}{\sqrt{-g}}\frac{\delta S_{m}}{\delta{g^{\mu \nu}}},
\ee is the energy-momentum tensor of matter. 

In the previous equations the spacetime metric $g$ does not contain any ultraviolet modification through the dependence with the energy of the particles probing the spacetime \cite{Magueijo:2002xx}. To study gravity's rainbow the spacetime metric $g$ must be replaced by the effective energy-dependent
rainbow metric  \cite{Magueijo:2002xx}
\be
\tilde{g}(\mathcal{E})=\eta_{A B} \tilde{e}^{A}(\mathcal{E})\otimes \tilde{e}^{B}(\mathcal{E}),
\label{rainbowg}
\ee where $\eta_{A B}=\text{diag}\,(-1,1,1,1)$ is the Minkowski tangent space metric, and $\tilde{e}^{A}$ are the corresponding energy-dependent tetrad frame such that 
\be
\tilde{e}^{\hat{0}}=\frac{1}{\tilde{f}(\mathcal{E})} \bar{e}^{\hat{0}},\:\:\: \tilde{e}^{\hat{i}}=\frac{1}{\tilde{g}(\mathcal{E})} \bar{e}^{\hat{i}},
\label{rainbowe}
\ee where  $\tilde{f}(\mathcal{E})$ and $\tilde{g}(\mathcal{E})$ are functions of the energy of the probe particle and $\bar{e}^{A}$ are the original frame fields without rainbow effect. According to the correspondence principle \cite{Magueijo:2002xx}, the rainbow functions $\tilde{f}(\mathcal{E})$ and $\tilde{g}(\mathcal{E})$ satisfy the limit conditions $\tilde{f}(\mathcal{E})\rightarrow{1}$ and $\tilde{g}(\mathcal{E})\rightarrow{1}$ for $\mathcal{E}/\mathcal{E}_{Pl}\ll 1 $, with $\mathcal{E}_{Pl}$ the Planck energy.
 Thus, in order to study rainbow gravity effects on the dynamics of inflation in $f(R)$ gravity, the spacetime metric consistent with homogeneity and isotropy conditions is assumed to be \cite{Ling:2006az}
\bea
ds^2=-\frac{\mathcal{N}^2}{\tilde{f}^2} dt^2+\frac{a^2}{\tilde{g}^2} \delta_{i j}dx^{i} dx^{j}.
\label{Back}
\eea At the limit $\tilde{f}, \tilde{g}\rightarrow 1$ we recover the standard Friedmann-Lemaître-Robertson-Walker (FLRW) background \cite{Weinberg:2008zzc}. Through the rescaling of the fields (i.e. conformal transformation) and field redefinitions one is led to the Einstein frame where the action of $f(R)$ gravity (Eq. \eqref{action1}) is seen to be equivalent to the action of a scalar field (inflaton) minimally coupled to gravity and directly coupled to matter \cite{DeFelice:2010aj}. During inflation the evolution of the universe is dominated by the dynamics of the inflaton and then the matter fluid can be neglected \cite{Weinberg:2008zzc}. Thus, the inflaton field propagating the degree of freedom from modified gravity in the field representation can be considered as our probe particle in the context of gravity's rainbow \cite{Ling:2006az, Chatrabhuti:2015mws}. 
Its energy density at the slow-roll regime is roughly proportional to the effective scalar potential, which ultimately becomes a function of the curvature scalar in the Jordan frame and then also of the Hubble expansion rate $H \equiv \frac{1}{a}\frac{da}{dt}$ \cite{DeFelice:2010aj}. Therefore, for an expanding universe the rainbow functions $\tilde{f}$, $\tilde{g}$, can be assumed to be functions of time \cite{Ling:2006az,Chatrabhuti:2015mws}.

Thus, the gravitational action of $f(R)$ gravity's rainbow becomes
\be
S=\frac{1}{2\kappa^2}\int{dtd^{3}{x}\left[\frac{a^3 \mathcal{N} f(R)}{\tilde{f}\tilde{g}^3}\right]} +S_m,
\label{Effective_action}
\ee where
\bea
&& R=\frac{6 \tilde{f}^2 (2 H^2+\dot{H})}{ \mathcal{N}^2}-\frac{6 H \tilde{f}^2  \dot{\mathcal{N}}}{\mathcal{N}^3}+ \frac{6 H \tilde{f} \dot{\tilde{f}}}{\mathcal{N}^2}-\nonumber\\
&& \frac{24 H\tilde{f}^2 \dot{\tilde{g}}}{\mathcal{N}^2 \tilde{g}}+\frac{6 \tilde{f}^2 \dot{\mathcal{N}} \dot{\tilde{g}}}{\mathcal{N}^3 \tilde{g}}-\frac{6 \tilde{f} \dot{\tilde{f}} \dot{\tilde{g}}}{\mathcal{N}^2 \tilde{g}}-\frac{6 \tilde{f}^2 \ddot{\tilde{g}}}{\mathcal{N}^2 \tilde{g}}+ \frac{18 \tilde{f}^2 \dot{\tilde{g}}^2}{\mathcal{N}^2 \tilde{g}^2}.
\eea
From the point of view of the effective action in Eq. \eqref{Effective_action}, one can treat $f(R)$ gravity's rainbow as an energy-dependent modified gravity theory \cite{He:2017mat}.  Thus, starting from action \eqref{Effective_action} we can study the slow-roll inflationary dynamics of the cosmological background and the evolution of the linear perturbations.

Varying the action with respect to $\mathcal{N}$ and taking $\mathcal{N}=1$ we obtain
\bea
&& 3 \left(H^2 F+H \dot{F}\right)-\frac{6 H F \dot{\tilde{g}}}{\tilde{g}}-\frac{3 \dot{F} \dot{\tilde{g}}}{\tilde{g}}+\frac{3 F \dot{\tilde{g}}^2}{\tilde{g}^2}-\frac{R F}{2 \tilde{f}^2}+ \frac{f}{2 \tilde{f}^2}=0,
\label{FR_00}
\eea while varying with respect to the scale factor $a$ yields
\bea
&& 3 H^2 F-6 H \dot{F}-\frac{3 \dot{F} \dot{\tilde{f}}}{\tilde{f}}-\frac{6 H F \dot{\tilde{g}}}{\tilde{g}}+\frac{6 \dot{F} \dot{\tilde{g}}}{\tilde{g}}+\frac{3 F \dot{\tilde{g}}^2}{\tilde{g}^2}-\nonumber\\
&& 3 \ddot{F}-\frac{3 f}{2 \tilde{f}^2}+\frac{R F}{2 \tilde{f}^2}=0.
\label{FR_ii}
\eea We have neglected the matter sector since the corresponding energy and pressure densities behave as inverse powers of the scale factor which grows quasi-exponentially during inflation. The set of equations \eqref{FR_00} and \eqref{FR_ii} constitute the background equations of $f(R)$ gravity's rainbow. In the next section we introduce the general setup for slow-inflation.

\section{General setup for slow-roll inflation}\label{Sec2}
In order to study slow-roll inflation we introduce the following set of parameters
\bea
&& \epsilon=-\frac{\dot{H}}{H^2},\:\:\:\delta_{\tilde{f}}=\frac{\dot{\tilde{f}}}{H \tilde{f}},\:\:\: \delta_{\tilde{g}}=\frac{\dot{\tilde{g}}}{H \tilde{g}},\nonumber\\
&& \eta_{\tilde{f}}=\frac{\dot{\delta}_{\tilde{f}}}{H \delta_{\tilde{f}}},\:\:\: \eta_{\tilde{g}}=\frac{\dot{\delta}_{\tilde{g}}}{H \delta_{\tilde{g}}}, \:\:\: \delta_{F}=\frac{\dot{F}}{H F},
\label{slow_para}
\eea During inflation the condition $\epsilon \ll 1$ is satisfied. To comply with this condition all the parameters defined in Eq.  \eqref{slow_para} are much smaller than the order of the unity. 

In terms of these parameters we can rewrite $R$ as

\bea
&& R=R_{0}\Bigg[1-\frac{1}{2}\epsilon+\frac{1}{2}\delta_{\tilde{f}}-2 \delta_{\tilde{g}}- \frac{1}{2}\delta_{\tilde{f}}\delta_{\tilde{g}}+  \delta_{\tilde{g}}^2+\frac{1}{2} \delta_{\tilde{g}}\epsilon-\frac{1}{2} \delta_{\tilde{g}}\eta_{\tilde{g}}\Bigg].
\label{R_Slow} 
\eea where $R_{0}= 12 H^2 \tilde{f}^2$. 
Thus, up to first order, Eq. \eqref{FR_00} is written as
\bea
&&2 \delta_{f_{,R}}-3 \mu \epsilon+ 3 \mu \delta_{\tilde{f}}+4 ( \mu-1)\delta_{\tilde{g}}+\mathcal{O}(\epsilon^2)=0,
\label{FR_00_Slow}
\eea where we have defined
\be
\delta_{f_{,R}}=\left.\frac{2 f-R F}{R F}\right|_{R=R_{0}},
\ee
\be
\mu \equiv \left.\frac{R F_{,R}}{F}\right|_{R=R_{0}}.
\ee and $\delta_{\mu}=\dot{\mu}/(H \mu)$.
Also, we have used the relation 
\be
\delta_{F}= 2 \mu (\delta_{\tilde{f}}-\epsilon)+\mathcal{O}(\epsilon^2).
\ee
Then, by solving Eq. \eqref{FR_00_Slow} for $\epsilon$ one obtains
\bea
\epsilon=\frac{2 \delta_{f_{,R}}}{3 \mu}+ \delta_{\tilde{f}}+\frac{4}{3}\delta_{\tilde{g}} \left(1-\frac{1}{\mu}\right),
\label{epsilon}
\eea and $\mu \neq 0$. 

Also, up to first order, Eq. \eqref{FR_ii} yields
\be
6\delta_{f,_{R}}- (4+5\mu) \epsilon+ (4 +5  \mu)\delta_{\tilde{f}}+12  (\mu -1) \delta_{\tilde{g}}+\mathcal{O}(\epsilon^2)=0.
\label{FR_ii_Slow}
\ee 

After substituting Eq. \eqref{epsilon} into \eqref{FR_ii_Slow}, one gets
\be
\left(1-\frac{1}{\mu}\right) \left[\delta_{f_{,R}}+2 \delta_{\tilde{g}} (\mu -1)\right]=0.
\ee 
This latter equation is satisfied for $\mu= 1+\mathcal{O}(\epsilon)$ and $\delta_{\mu}=\mathcal{O}(\epsilon^2)$. For instance, this holds for the kind of functions $f(R)=R+R^2 \mathcal{A}(R)$. The case $\mathcal{A}(R)=1/(6 M^2)$ with $M$ the mass scale corresponds to Starobinsky model \cite{Starobinsky}. In general $\mathcal{A}(R)$ has to be a slowly varying function which meets the slow-roll conditions $\left|\mathcal{A}'(R)\right|\ll \mathcal{A}(R)/R$ and $\left|\mathcal{A}''(R)\right|\ll \mathcal{A}(R)/R^2$ \cite{Ketov:2017aau,DeFelice:2010aj}. In fact,  for these models one finds $\mu=1-\delta_{f_{,R}}$ with $\delta_{f_{,R}}=1/(2 R_{0} \mathcal{A}(R_{0})) \ll 1 $. Notice that from Eqs. \eqref{FR_00_Slow} and \eqref{epsilon}, one finds that the parameter $\delta_{\tilde{g}}$ associated with the rainbow function $\tilde{g}$ does not contribute to the dynamics of slow-roll inflation at first-order.   


Below we study the evolution of the cosmological perturbations around the background metric \eqref{Back}.

\section{Primordial Fluctuations}\label{Sec3}

\subsection{Scalar Perturbations}
By starting from the general perturbed metric about the flat FLRW background \cite{DeFelice:2010aj}, and using Eqs. \eqref{rainbowg} and \eqref{rainbowe},  the corresponding perturbed metric with rainbow effect can be obtained. To study the primordial scalar fluctuations generated during inflation we assume the following scalarly perturbed metric with rainbow effect
\bea
&& ds^{2}=-\frac{1+2 \alpha}{\tilde{f}^2} dt^2-\frac{2 a}{\tilde{f} }\partial_{i}{\beta}dt dx^{i}+ a^2 \Big(\delta_{i j}+2 \psi \delta_{i j}+2 \partial_{i}\partial_{j}{\gamma}\Big)dx^{i}dx^{j}.
\label{Scalar_Pert_metric}
\eea For simplicity we assumed $\tilde{g}=1$. Thus, for $\tilde{f}=1$ one recovers the usual scalarly perturbed metric \cite{DeFelice:2010aj}. 

Also, for convenience we introduce the following perturbed variables
\be
\chi\equiv a \left(\beta+a \dot{\gamma}\right),\:\:\:\:\: A\equiv 3\left(H \alpha-\dot{\psi}\right)-\frac{\Delta}{a^{2}}\chi.
\ee

Then, the perturbed field equations with rainbow effects are given by
\bea
\label{Pert_Eq_00}
&& \frac{\Delta{\psi}}{a^2}+ H \tilde{f}^2 A+\frac{\tilde{f}\left(\tilde{f}-1\right)\Delta{\beta}}{a} \Bigg(H+\frac{\dot{F}}{2  F}\Bigg)=-\frac{1}{2 F}\Bigg[\Bigg(3 H^2 \tilde{f}^2+3 \tilde{f}^2 \dot{H}+3 H \tilde{f} \dot{\tilde{f}}+\frac{\Delta}{a^2}\Bigg)\delta{F}-\nonumber\\
&& 3 H \tilde{f}^2\delta{\dot{F}}+
3 H \tilde{f}^2 \dot{F}\alpha+\tilde{f}^2 \dot{F} A\Bigg],\\
\label{Pert_Eq_0i}
&& H \alpha-\dot{\psi}=-\frac{1}{2F}\Bigg[H \delta{F}+ \dot{F} \alpha-\delta{\dot{F}}\Bigg],\\
\label{Pert_Eq_i_neq_j}
&& \dot{\chi}+\left(H+\frac{\dot{\tilde{f}}}{\tilde{f}}\right) \chi-\frac{ \alpha}{\tilde{f}^2}-\frac{\psi }{\tilde{f}^2}+a \beta\Bigg(\frac{\dot{F}}{\tilde{f} F}-\frac{ \dot{F}}{F}+\frac{2  H}{\tilde{f}}-2  H-\frac{ \dot{\tilde{f}}}{\tilde{f}}\Bigg)-a \left(1-\frac{ 1}{\tilde{f}}\right) \dot{\beta}\nonumber\\
&& =\frac{1}{F}\Bigg(\frac{ \delta{F}}{\tilde{f}^2}-\dot{F} \chi\Bigg),\\
\label{Pert_Eq00_Trace}
&& \dot{A}+\left(2 H+\frac{\dot{\tilde{f}}}{\tilde{f}}\right) A+\left(3\dot{H}+\frac{\Delta}{a^2 \tilde{f}^2} \right)\alpha+\frac{\Delta{\beta}}{a}\Bigg(H-\frac{H }{\tilde{f}}- \frac{ \dot{F}}{2 \tilde{f} F}+\frac{\dot{F}}{2 F}+\frac{\dot{\tilde{f}}}{ \tilde{f}}\Bigg)+\nonumber\\
&& \left(1-\frac{1}{\tilde{f}}\right) \frac{\Delta{\dot{\beta}}}{a} =\frac{1}{2 F}\Bigg[3 \delta{\ddot{F}}+3\left(H+\frac{\dot{\tilde{f}}}{\tilde{f}}\right)\delta{\dot{F}}-6 \Bigg(H^2+\frac{\Delta}{6 a^2 \tilde{f}^2}\Bigg) \delta {F}-3 \dot{F}\dot{\alpha}-\dot{F}A-\nonumber\\
&& 3\Bigg(H\dot{F}+2 \ddot{F}+\frac{2 \dot{F} \dot{\tilde{f}}}{\tilde{f}}+\frac{2 H F \dot{\tilde{f}}}{\tilde{f}}\Bigg)\alpha\Bigg],
\\
&& \delta{\ddot{F}}+\left(3 H+\frac{\dot{\tilde{f}}}{\tilde{f}}\right) \delta{\dot{F}}-\left(\frac{\Delta}{a^2}+\frac{R}{3 }\right)\frac{\delta{F}}{\tilde{f}^2}+\frac{\dot{F}}{a}\left(\frac{1}{ \tilde{f}}-1\right)\Delta{\beta}=
\dot{F}\left(A+\dot{\alpha}\right)+\nonumber\\
&& \Bigg(2\ddot{F}+3 H\dot{F}+\frac{2\dot{F}\dot{\tilde{f}}}{\tilde{f}}\Bigg)\alpha-\frac{F \delta{R}}{3 \tilde{f}^2},
\eea where $\delta{R}$ is given by
\bea
&& \delta{R}=-2 \Bigg[\tilde{f}^2 \dot{A}+A\left(4 H \tilde{f}^2+  \tilde{f} \dot{\tilde{f}}\right)+ \Bigg(\frac{\Delta}{a^2}+3\tilde{f}^2 \dot{H}+3 H \tilde{f} \dot{\tilde{f}}\Bigg)\alpha+\frac{2   \Delta{\psi}}{a^2}+\nonumber\\
&& \frac{\Delta{\beta}}{a} \Bigg(3 H \tilde{f}^2-3 H \tilde{f} +\tilde{f} \dot{\tilde{f}}\Bigg)+\left(\tilde{f}- 1\right)\frac{\tilde{f}\Delta{\dot{\beta}}}{a}\Bigg].
\eea

Now let us consider the gauge transformation \cite{DeFelice:2010aj}
\be
\hat{t}=t+\delta{t},\:\:\: \hat{x}^{i}=x^{i}+\delta^{i j}\partial_{j}{\delta{x}}.
\ee Then the scalar perturbations transform as
\bea
&& \hat{\alpha}=\alpha+\frac{\dot{\tilde{f}}}{\tilde{f}}\delta{t}-\dot{\delta{t}},\\
&& \hat{\beta}=\beta-\frac{1}{a \tilde{f}}\delta{t}+a \tilde{f}\dot{\delta{x}},\\
&& \hat{\psi}=\psi-H\delta{t},\\
&& \hat{\gamma}=\gamma-\delta{x},
\eea and 
also
\be
\hat{\delta{F}}=\delta{F}-\dot{F}\delta{t}.
\ee
In this way we define the following gauge invariant variables
\bea
\label{InvPhi}
&& \Phi=\alpha -\tilde{f}\frac{d}{dt}\left[a^2\tilde{f}\left(\dot{\gamma}+\frac{\beta}{a \tilde{f}}\right)\right],\\
\label{InvPsi}
&& \Psi=-\psi+a^2 \tilde{f}^2 H\left(\dot{\gamma}+\frac{\beta}{a \tilde{f}}\right),\\
&& \mathcal{R}=\psi-\frac{H}{\dot{F}}\delta{F}.
\label{mathR}
\eea 

By using the above set of perturbed field equations one can obtain the evolution equation for the primordial curvature perturbation.  In a $4D$ diffeomorphism invariant theory, one can fix, at most, two scalar perturbation fields, either $\gamma=0=\beta$ (Newtonian gauge)
or $\gamma=0=\delta{F}$ (uniform field gauge) \cite{DeFelice:2010aj}. Thus, from the gauge-invariant quantities defined in Eqs.  \eqref{InvPhi}, \eqref{InvPsi} and \eqref{mathR}, one can completely fix the gauge degree of freedom by choosing the specific gauge conditions $\gamma=0$ and $\delta{F}=0$ \cite{DeFelice:2010aj}. Notice that we have assumed in Eq. \eqref{mathR} that $\dot{F}=F_{,R} \dot{R}\neq 0$ which is valid during inflation for a quasi-de Sitter background. Likewise, we have neglected the matter perturbations \cite{DeFelice:2010aj}.

\subsubsection{Mukhanov-Sasaki equation}
We assume the gauge conditions  $\gamma=0$ and $\delta{F}=0$ \cite{DeFelice:2010aj}. In this case one has $\mathcal{R}=\psi$. Thus, from Eq. \eqref{Pert_Eq_0i} we obtain
\be
\alpha= \frac{ \dot{\mathcal{R}}}{H+\frac{\dot{F}}{2 F}}.
\label{Eq_Phi}
\ee Then, by replacing the previous result into Eq. \eqref{Pert_Eq_00} one is led to
\bea
&& A=- \frac{1}{H+\frac{\dot{F}}{2 F}}\Bigg[\frac{1}{a^2 \tilde{f}^2} \Delta{\mathcal{R}} + \frac{3 H\dot{F}}{2 F\left( H+\frac{\dot{F}}{2 F}\right)}\dot{\mathcal{R}}\Bigg]-\frac{(\tilde{f}-1) \Delta{\beta}}{a \tilde{f}}.
\label{Eq_A}
\eea Also, by putting the background equations into \eqref{Pert_Eq00_Trace} one finds
\bea
&& \dot{A}+\left(2 H+\frac{\dot{F}}{2 F}+\frac{\dot{\tilde{f}}}{\tilde{f}}\right)A+\frac{3\dot{F}}{2 F}\dot{\alpha}+\Bigg(\frac{3\ddot{F}}{2 F}+\frac{3 H \dot{F}}{F}+\frac{3\dot{F}\dot{\tilde{f}}}{2 \tilde{f}F}\Bigg)\alpha+\nonumber\\
&& \frac{\Delta{\alpha}}{a^2 \tilde{f}^2}+\left(H-\frac{H}{\tilde{f}}-\frac{\dot{F}}{2\tilde{f} F}+\frac{\dot{F}}{2 F}+\frac{\dot{\tilde{f}}}{\tilde{f}}\right)\frac{\Delta{\beta}}{a}+\left(1-\frac{1}{\tilde{f}}\right) \frac{\Delta{\dot{\beta}}}{a}=0.
\label{Eq_dotA}
\eea Finally, by combining Eqs. \eqref{Eq_Phi}, \eqref{Eq_A}, \eqref{Eq_dotA}, and with the help of the background equations, one finds in Fourier space the following equation for $\mathcal{R}$
\be
\ddot{\mathcal{R}}+\frac{\left(a^3 Q_{s}\right)^{\cdot}}{a^3 Q_{s}}\dot{\mathcal{R}}+\frac{c_{s}^2 k^2}{a^2}\mathcal{R}=0,
\label{Eq_R}
\ee where 
\be
Q_{s}=\frac{3\tilde{f}\dot{F}^2}{2 \kappa^2 F \left(\frac{\dot{F}}{2 F}+H\right)^2},
\ee and
\be
c_{s}^2=\frac{1}{\tilde{f}^2}.
\ee
At this point it is useful to introduce the parameters
\bea
&& \eta\equiv \frac{\dot{Q}_{s}}{H Q_{s}}=3 \delta_{\tilde{f}}-2 \epsilon+2 \eta_{F},\nonumber \\
&& =3 \delta_{\tilde{f}}-2 \epsilon+2 \delta_{\mu}+\frac{2 \delta_{\tilde{f}} \eta_{\tilde{f}}}{\delta_{\tilde{f}}-\epsilon}- \frac{2 \eta_{\epsilon} \epsilon}{\delta_{\tilde{f}}-\epsilon},\\
&& s\equiv \frac{\dot{c}_{s}}{H c_{s}}=-\delta_{\tilde{f}},
\eea where we also defined
\be
\eta_{F}=\frac{\dot{\delta}_{F}}{H\delta_{F}},\:\:\:\eta_{\epsilon}=\frac{\dot{\epsilon}}{H \epsilon}.
\ee Furthermore, we have used the relation
\be
\eta_{F}=\delta_{\mu}+ \frac{\delta_{\tilde{f}} \eta_{\tilde{f}}}{\delta_{\tilde{f}}-\epsilon}-\frac{\eta_{\epsilon} \epsilon}{\delta_{\tilde{f}}-\epsilon}.
\ee
Also, the conditions for the absence of ghost and Laplacian instabilities read as $Q_{s}>0$ and $c_{s}^2>0$. The former condition implies $\tilde{f}/F>0$. In particular, for $\tilde{f}>0$ one requires $F=f_{,R}>0$.

Notice that the propagation speed of the scalar perturbations $c_{s}(t)$ is time dependent due to the rainbow effects. Of course, this is not unusual since in rainbow gravity the speed of the light is time dependent \cite{Magueijo:2002xx}. Thus, in order to obtain the propagation speed of the scalar modes equal to $1$ (speed of light in natural units), we move to the ``Rainbow Frame" where a new unit of time (``sound-horizon" time) is assumed. So, we replace the 
standard conformal time $d\tau=a^{-1} dt$ by the sound-horizon time $d\tau_{\text{RF}}=c_{s}(\tau) d\tau=(c_{s}(t)/a)dt$ \cite{Amelino-Camelia:2013wha}. Using the canonically-normalized Mukhanov variable $v=z \mathcal{R}$ and $z^2=2 a^2 Q_{s} c_{s}$ the equation \eqref{Eq_R} is written as 
\be
v''_{k}+(k^2+M^2)v_{k}=0,
\label{MS}
\ee where
\be
M^{2}=-\frac{z''}{z}=-\left(\frac{a H}{c_{s}}\right)^2 \left[2-\epsilon+\frac{3}{2}\eta+\frac{1}{2}s\right].
\ee

Thus, we write Eq. \eqref{MS} in the following form
\be
v''_{k}+\left[k^2-\frac{1}{\tau_{\text{RF}}^2}\left(\nu^2-\frac{1}{4}\right)\right] v_{k}= 0, 
\label{MS2}
\ee where 
\be
\nu^2=\frac{9}{4}+3\epsilon +\frac{3}{2}\eta+\frac{9}{2}s.
\ee 

The general solution to Eq.\eqref{MS2} is
\be
v_{k}(\tau_{\text{RF}})=\sqrt{-\tau_{\text{RF}}}\left[C_{1} H_{\nu}^{(1)}(-k \tau_{\text{RF}})+C_{2} H_{\nu}^{(2)}(-k \tau_{\text{RF}})\right], 
\ee being $H_{\nu}^{(1,2)}$ the Hankel’s functions of first and second kind, respectively \cite{Riotto:2018pcx}. Assuming the Bunch-Davies vacuum $v_{k}(\tau_{\text{RF}})\simeq e^{-i k \tau_{\text{RF}}}/\sqrt{2 k}$ at the ultraviolet regime $-k \tau_{\text{RF}}\gg 1$, and using some identities, one obtains
\be
v_{k}(\tau_{\text{RF}})=\frac{\pi}{2} e^{i \frac{\pi}{2}(\nu+\frac{1}{2})} \sqrt{-\tau_{\text{RF}}} H_{\nu}^{(1)}(-k \tau_{\text{RF}}).
\ee On super-horizon scales $-k \tau_{\text{RF}}\ll 1$ one gets \be
v_{k}(\tau_{\text{RF}})=\frac{2^{\nu-\frac{3}{2}}}{\sqrt{2 k}}e^{i\frac{\pi}{2}(\nu-\frac{1}{2})} \frac{\Gamma(\nu)}{\Gamma(\frac{3}{2})} (-k \tau_{\text{RF}})^{\frac{1}{2}-\nu},
\ee and then 
\bea
&& \left|R_{k}\right|=z^{-1}\left|v_{k}\right|\simeq \frac{H}{2 \sqrt{Q_{s}c_{s}^3 k^3}}\left(\frac{\tau_{\text{RF}}}{\tau_{\text{RF}}^{*}}\right)^{\frac{3}{2}-\nu},\nonumber\\
&& \simeq \frac{H_{*}}{2 \sqrt{Q_{s*}c_{s*}^{3} k^3}}.
\label{Rk1}
\eea In this latter equation one has that $\tau_{\text{RF}}^{*}\simeq -1/k$ is the value of $\tau_{\text{RF}}$ at the horizon crossing. Also, we used $H\simeq H_{*} \left(\tau_{\text{RF}}/\tau_{\text{RF}}^{*}\right)^{\epsilon}$,  $Q_{s}\simeq Q_{s*} \left(\tau_{\text{RF}}/\tau_{\text{RF}}^{*}\right)^{-\eta}$, and $c_{s}\simeq c_{s*}\left(\tau_{\text{RF}}/\tau_{\text{RF}}^{*}\right)^{-s}$ where $H_{*}$, $Q_{s*}$ and $c_{s*}$ are evaluated at $\tau_{\text{RF}}=\tau_{\text{RF}}^{*}$ \cite{Leyva:2021fuo}.

Therefore, the scalar power spectrum of curvature fluctuation is given by
\bea
&& \mathcal{P}_{s}(k)\equiv \frac{k^3}{2 \pi^2}\left|\mathcal{R}_{k}\right|^2
\simeq \frac{H_{*}^2}{8 \pi^2 Q_{s*} c_{s*}^3}.
\label{Ps}
\eea

The scalar spectral index is
\bea
&& n_{s}-1\equiv \frac{d \ln{\mathcal{P}_{s}(k)}}{d \ln{k}}\simeq -2 \epsilon-\eta-3 s,\nonumber\\
&& \simeq - 2 \eta_{f_{,R}}.
\label{ns}
\eea The slow-roll parameter $\delta_{\tilde{f}}$ related to gravity's rainbow does not explicitly contribute to $n_s$. However, these contributions arise implicitly into the term $\eta_{f_{,R}} =\dot{\delta}_{f_{,R}}/(H\delta_{f_{,R}})$.

Below we study tensor perturbations. 


\subsection{Tensor Perturbations}

For tensor perturbations one writes the perturbed FLRW metric with rainbow effect as follows 
\be 
ds^2=-\frac{1}{\tilde{f}^2}dt^2+a^2 \left(\delta_{i j}+h_{i j}\right).
\ee where $h_{i j}$ are the tensor perturbations. These tensor perturbations $h_{i j}$ can be decomposed in terms of their two polarization states $h_{+}$, $h_{\times}$, such that \cite{DeFelice:2010aj}
\be
h_{i j}=h_{+}e^{+}_{i j}+h_{\times}e^{\times}_{i j},
\ee where $e^{+}_{i j}$, $e^{\times}_{i j}$ are the corresponding polarization tensors. 

By substituting this perturbed metric into the modified field equations \eqref{FEQ} one obtains
\be
\ddot{h}_{\theta}+\frac{\left(a^3 Q_{t}\right)^{\cdot}}{a^3 Q_{t}}\dot{h}_{\theta}+\frac{c_{t}^2 k^2}{a^2} h_{\theta}=0,
\label{Eq.h}
\ee with $\theta=+, \times$, and where 
\be
Q_{t}=\frac{F \tilde{f}}{2 \kappa^2},
\ee and
\be
c_{t}^2=\frac{1}{\tilde{f}^2}. 
\ee  We define
\bea
&& \eta_{t}\equiv \frac{\dot{Q}_{t}}{H Q_{t}}=  \delta_{\tilde{f}}+\delta_{F},\nonumber\\
&& = \delta_{\tilde{f}}+2 (\delta_{\tilde{f}}-\epsilon)\mu.
\eea and 
\be
s_{t}\equiv \frac{\dot{c_{t}}}{H c_{t}}=s.
\ee
Following a similar procedure than in the case of scalar perturbations, one can obtain the primordial power spectrum for tensor perturbations \cite{DeFelice:2010aj}. 

\subsubsection{Mukhanov-Sasaki equation}

Now we introduce the canonically-normalized field $v_{\theta}=z_{t} h_{\theta}$ where $z_{t}^2=2 a^2 Q_{t} c_{t}$, along with the sound-horizon time $d\tau_{\text{RF}}=c_{t}(\tau) d\tau=(c_{t}(t)/a)dt$ \cite{Amelino-Camelia:2013wha}. Then, Eq. \eqref{Eq.h} is turned into 
\be
v_{\theta,k}''+\left(k^2+M_{t}^2\right) v_{\theta,k}=0, 
\ee
where
\be
M_{t}^2=-\frac{z_{t}''}{z_{t}}=-\left(\frac{a H}{c_{t}}\right)^2\left[2-\epsilon +\frac{3}{2}\eta_{t}+\frac{1}{2} s_{t}\right].
\ee Furthermore this latter equation can be written as 
\be
v''_{\theta,k}+\left[k^2-\frac{1}{\tau_{\text{RF}}^2}\left(\nu_{t}^2-\frac{1}{4}\right)\right] v_{\theta,k}= 0, 
\label{MS2_t}
\ee where 
\be
\nu_{t}^2=\frac{9}{4}+3\epsilon +\frac{3}{2}\eta_{t}+\frac{9}{2}s_{t}.
\ee Following a similar procedure to what was performed in the case of scalar perturbations, one obtains
\be
\mathcal{P}_{t}=\frac{H_{*}^2}{2 \pi^2 Q_{t*}c_{t*}^3}.
\ee 
Thus, the tensor spectral index is 
\bea
&& n_{t}\equiv \frac{d \ln{\mathcal{P}_{t}}}{d \ln{k}}=-2 \epsilon-\eta_{t}-3 s_{t},\nonumber\\
&& =\frac{4}{3}\delta_{f_{,R}} \left(1-\frac{1}{ \mu}\right).
\label{nT}
\eea For $\mu=1+\mathcal{O}(\epsilon)$ one obtains 
\be
n_{t}=0.
\ee

The tensor-to-scalar ratio is calculated as
\bea
&& r =\frac{\mathcal{P}_{t}}{\mathcal{P}_{s}}=12 \delta_{F}^2=48 \left(\delta_{\tilde{f}}-\epsilon\right)^2,\nonumber\\
&& =\frac{64}{3 \mu^2}\delta_{f_{,R}}^2. 
\label{r}
\eea For $\mu=1+\mathcal{O}{(\epsilon)}$ one obtains
\be
r=\frac{64}{3}\delta_{f_{,R}}^2.
\ee The contributions coming from gravity's rainbow appear implicitly into the slow-roll parameter $\delta_{f_{,R}}$.

\section{Starobinsky inflation with rainbow gravity effects}\label{Sec4}

In this section we assume \cite{Starobinsky}
\be
f(R)=R+\frac{R^2}{6 M^2},
\ee where $M$ is the mass-energy scale.
Also, we consider \cite{Chatrabhuti:2015mws}
\be
\tilde{f}=\left(\frac{H}{M}\right)^{\lambda}.
\ee  Notice that during inflation $H\gtrsim M$ \cite{DeFelice:2010aj}. Then, one obtains
\bea
&& \delta_{f_{,R}}\simeq \frac{1}{4} \left(\frac{M}{H}\right)^{2\left(1+ \lambda\right)},\\
&& \eta_{f_{,R}}\simeq 2(1+\lambda) \epsilon,\\
&& \delta_{\tilde{f}}\simeq -\lambda \epsilon.
\eea From Eq. \eqref{epsilon} we get
\be
\epsilon\simeq \frac{1}{6 (\lambda +1)}\left(\frac{M}{H}\right)^{2\left( \lambda+1\right)}.
\ee From the latter equation one obtains
\be
H\simeq H_{i}-\frac{M^2}{6 (1+\lambda)}\left(\frac{M}{H_{i}}\right)^{2 \lambda}\left(t-t_{i}\right),
\label{H_t}
    \ee where $H_{i}=H(t_{i})$ is the value of the Hubble parameter at the beginning of inflation. For $\lambda=0$, the equation \eqref{H_t} gives us the standard solution found in Starobinsky inflation \cite{DeFelice:2010aj}. Also, for $\lambda \neq 0$, this equation is consistent with what was obtained in Ref. \cite{Chatrabhuti:2015mws}.  The number of e-folding of inflation is 
\bea
&& N=\int_{t_{i}}^{t_{f}}{H dt}\simeq H_{i}\left(t_{f}-t_{i}\right)- \frac{M^2}{12 \left(1+\lambda\right)}\left(\frac{M}{H_{i}}\right)^{2 \lambda}\left(t_{f}-t_{i}\right)^2.
\eea At the end of inflation $\epsilon (t_{f}) \simeq 1$ and then $t_{f}\simeq t_{i}+\frac{6\left(1+\lambda\right)}{M}\left(\frac{H_{i}}{M}\right)^{1+2 \lambda}$. Therefore
\be
\left(\frac{H_{i}}{M}\right)^{2\left(1+\lambda\right)} \simeq  \frac{N}{3\left(1+\lambda\right)}. 
\ee At the time of the horizon crossing, $t=t_{k}$,  one has $H_{k}\equiv H(t_{k})\simeq H_{i}$ and then
\be
\epsilon\simeq  \frac{1}{2 N_{k}}.
\ee Thus, we obtain
\bea
&& \delta_{f_{,R}}\simeq \frac{3\left(1+\lambda\right)}{4 N_{k}},\\
&& \eta_{f_{,R}}\simeq \frac{1+\lambda}{N_{k}},\\
&& \delta_{\tilde{f}}\simeq -\frac{\lambda}{2 N_{k}}.
\eea Therefore, the scalar power spectrum yields
\be
\mathcal{P}_{s}=\frac{1}{48 \pi ^2}\left(\frac{M}{M_{pl}}\right)^2\frac{N_{k}^2}{(1+\lambda)^2}.
\ee From the latest Planck data the amplitude of primordial scalar perturbations is $\mathcal{P}_{s}=2.141\times 10^{-9}$ \cite{Planck:2018jri}. Using this latter equation one can constrain the mass scale $M$ after constraining the $\lambda$ parameter on the $n_{s}-r$ plane.

The scalar spectral index is
\be
n_{s}\simeq 1-\frac{2\left(1+\lambda\right)}{N_{k}}.
\label{ns_rm}
\ee

The tensor-to-scalar ratio becomes
\be
r\simeq \frac{12 \left(1+\lambda\right)^2}{N_{k}^2}.
\ee 
The latest cosmological data from Planck satellite \cite{Planck:2018jri} give the following constraint for the scalar spectral index
\be
n_{s}=0.9649 \pm 0.0042,
\label{n_s_C}
\ee at $68 \%$ C.L. On the other hand, new data from BICEP/ Keck XIII have recently been published \cite{BICEPKeck:2021gln} which put stronger constraints on the amplitude of primordial gravitational waves and then on the tensor-to-scalar ratio
\be
r<0.036,
\ee at $95 \%$ C.L. Using these observational constraints on $n_{s}$ and $r$ we obtain
\be
50.89<N_{k}\leq 64.72,\:\:\: \text{and}\:\:\:0<\lambda <0.01965 N_{k}-1,
\label{Constraint1}
\ee or
\bea
&& N_{k}>64.72\:\:\:\text{and}\nonumber\\
&& 0.01545 N_{k}-1<\lambda <0.01965 N_{k}-1.
\label{Constraint2}
\eea For instance, for $N_{k}=60$ one finds
\be
0<\lambda <0.179,
\ee and then
\be
1.6785<\frac{M}{M_{pl}}\times 10^{5}<1.9790.
\ee

while for  $N_{k}=65$
\be
0.00425<\lambda <0.27725,
\ee and then
\be
1.55599<\frac{M}{M_{pl}}\times 10^{5}<1.97898.
\ee

\begin{figure}[htbp]
	\centering	\includegraphics[width=0.55\textwidth]{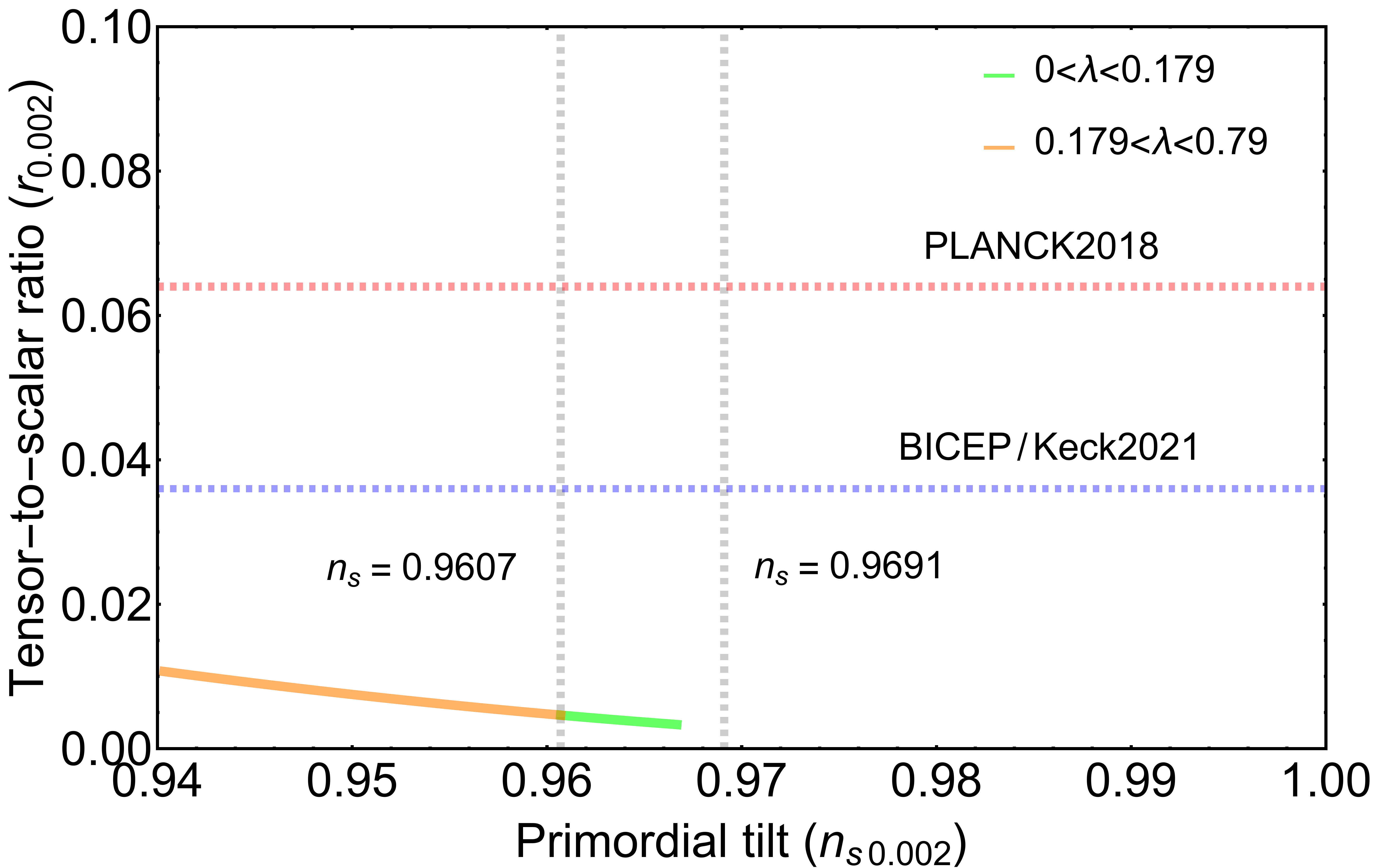}
	\caption{ To confront with the predictions of the model we used both the PLANCK $2018$ data \cite{Planck:2018jri} and the recently released BICEP/Keck data \cite{BICEPKeck:2021gln}. We used $N_k=60$.}  
	\label{FIG1}
\end{figure}

\begin{figure}[htbp]
	\centering	\includegraphics[width=0.55\textwidth]{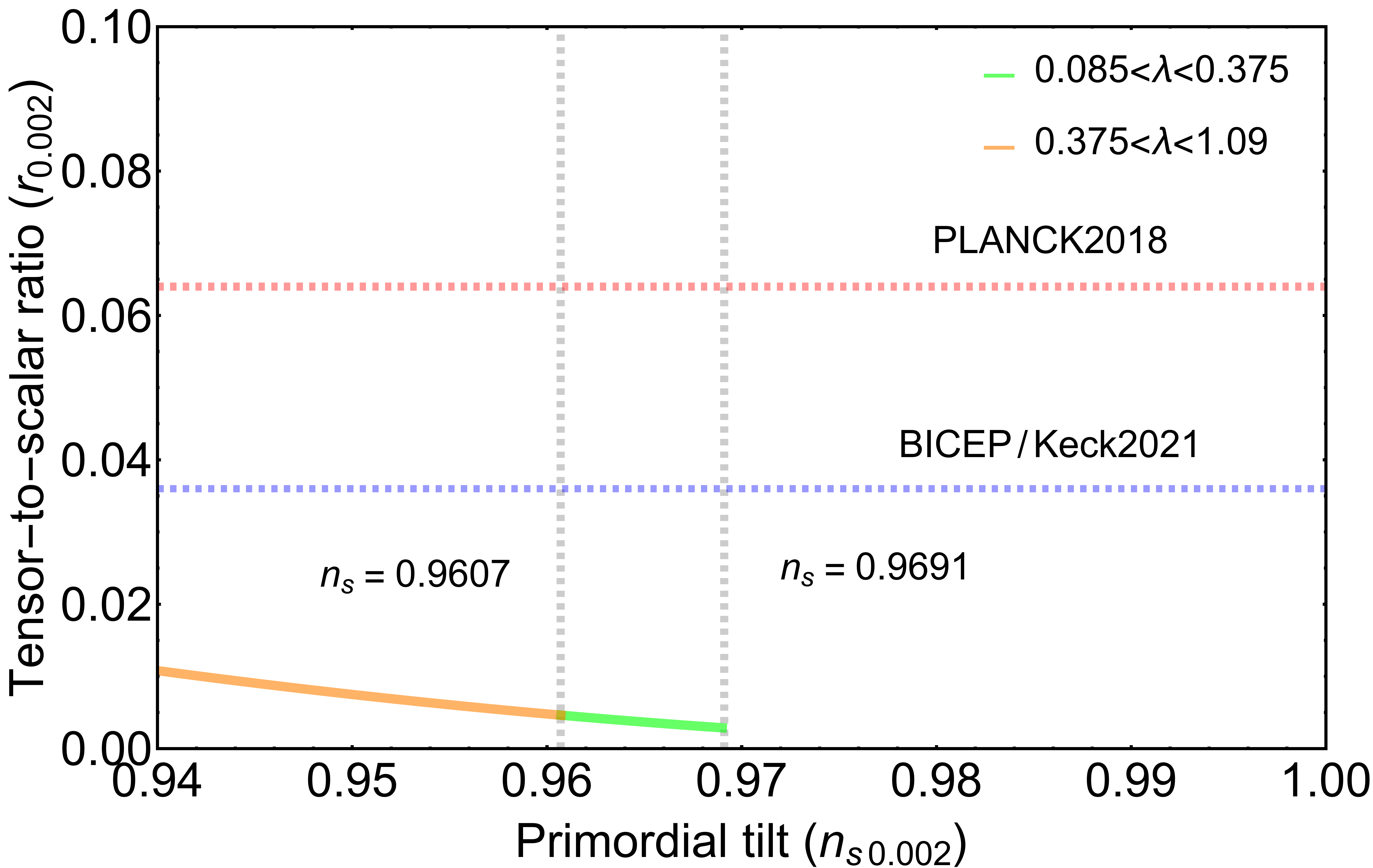}
	\caption{ To confront with the predictions of the model we used both the PLANCK $2018$ data \cite{Planck:2018jri} and the recently released BICEP/Keck data \cite{BICEPKeck:2021gln}.  We used $N_k=70$.}  
	\label{FIG2}
\end{figure}

\begin{figure}[htbp]
	\centering	\includegraphics[width=0.55\textwidth]{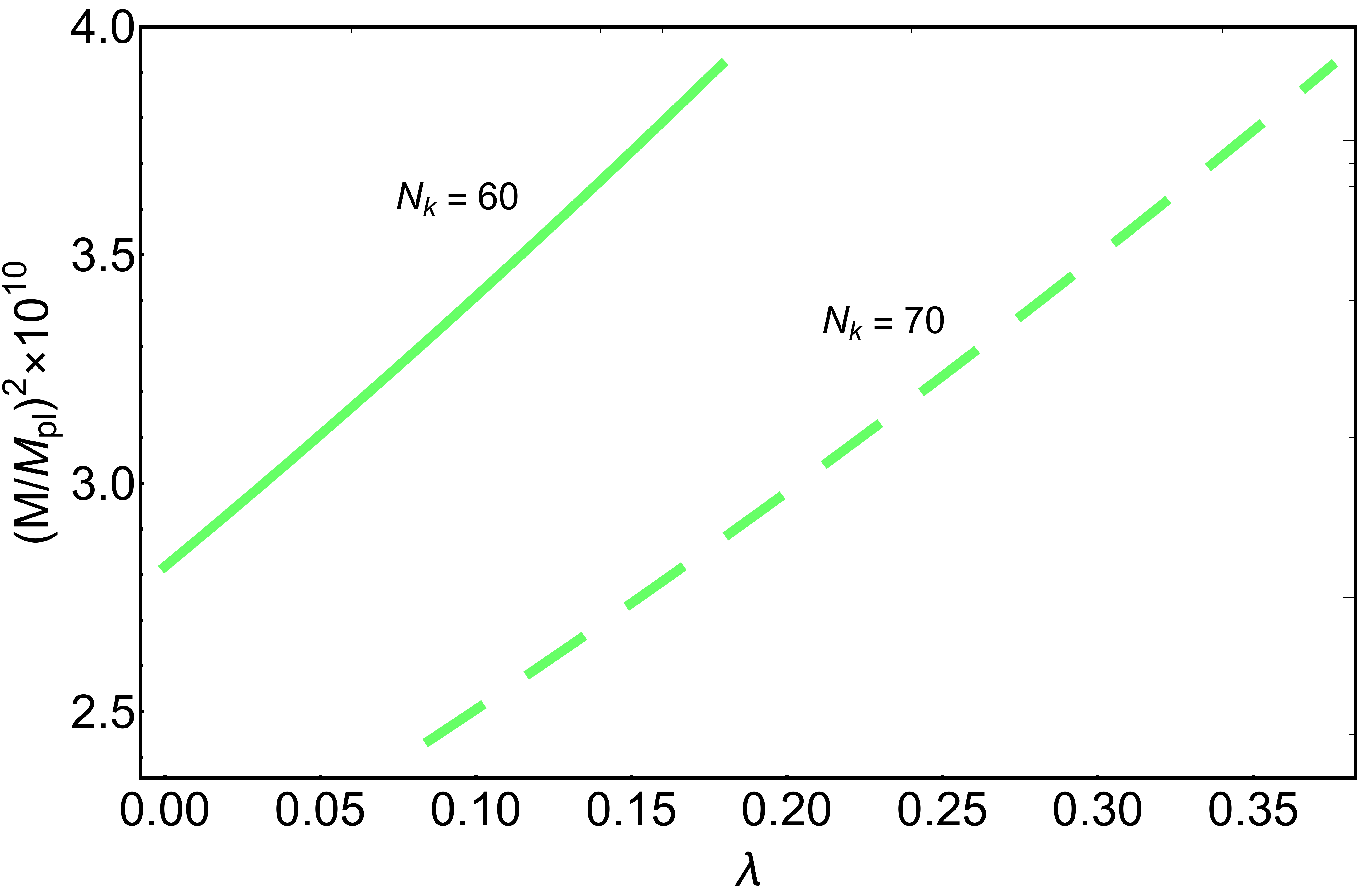}
	\caption{We depict the behavior of the mass scale $M$ as a function of the rainbow parameter $\lambda$ for several different values of the $e$-folding number of inflation $N_k$.}  
	\label{FIG3}
\end{figure}

\begin{figure}[htbp]
	\centering	\includegraphics[width=0.55\textwidth]{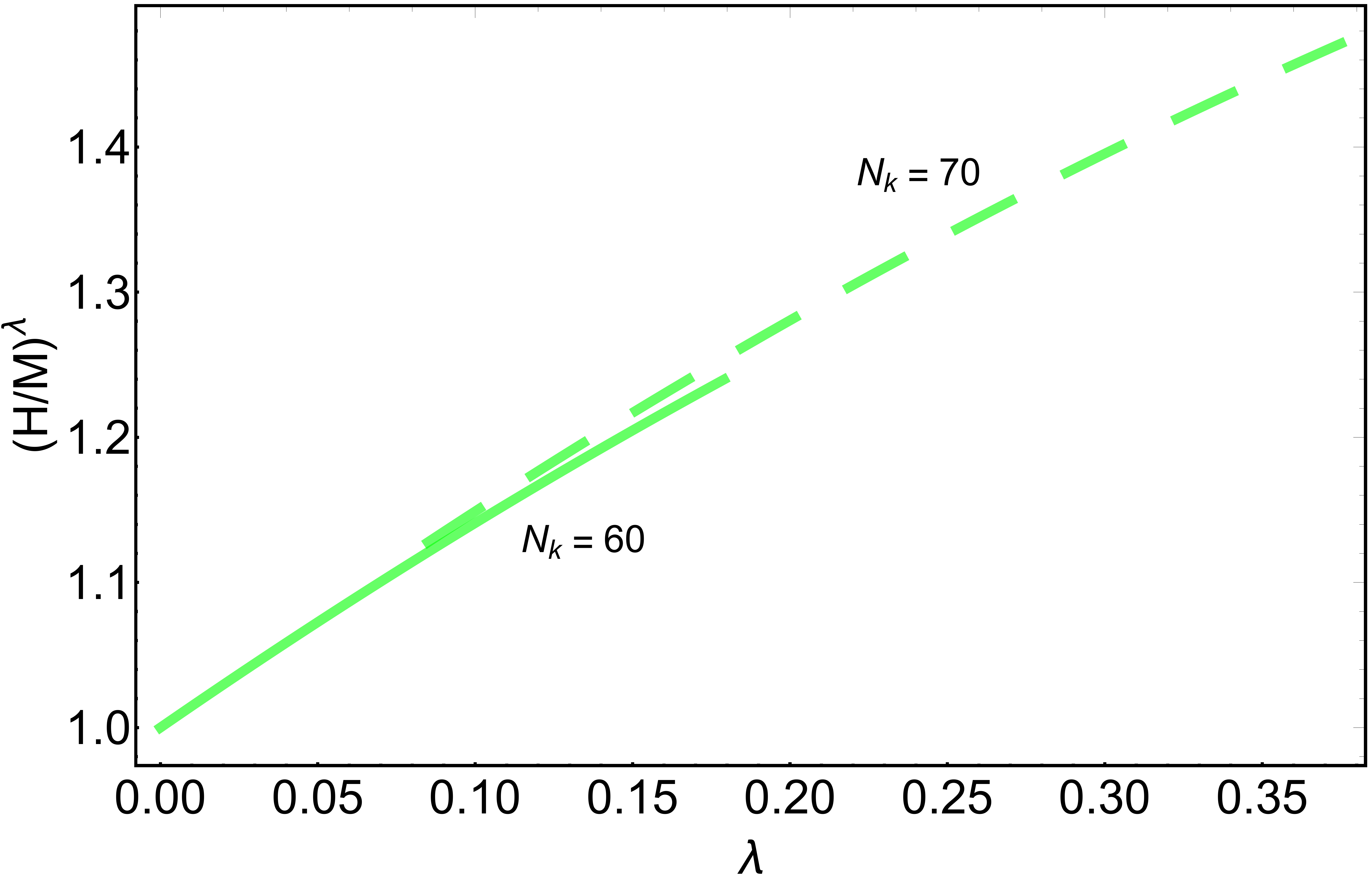}
	\caption{We show the behavior of the rainbow function $\tilde{f}=(H/M)^\lambda$ in terms of the parameter $\lambda$ for several  different values of the $e$-folding number of inflation $N_k$.}  
	\label{FIG4}
\end{figure}

In FIGS \ref{FIG1} and \ref{FIG2}, we depict the predictions of the model in the $n_{s}-r$ plane. We have used the latest data from Planck 2018 and BICEP/Keck 2021. These results are consistent with the constraints found for $\lambda$ in Eqs. \eqref{Constraint1} and \eqref{Constraint2}. Particularly, one can see that small values of $\lambda$ ($\lambda<1$) are preferred. Higher values of $\lambda$ can also give results compatible with observations, but this would require higher values of the $e$-folding number of inflation $N_k$. For instance, for $\lambda\gtrsim 1$, the values $N_{k} \gtrsim 110$ would be required. Otherwise, higher values of $\lambda$ could violate the slow-roll conditions. Also, this could conflict with the expected results from the subsequent reheating era after inflation \cite{Kofman:1994rk,Kofman:1997yn,Bassett:2005xm,Lopez:2021agu}. In FIG \ref{FIG3} we show the behavior of the mass scale $M$ as a function of the rainbow parameter $\lambda$ for different values of $N_k$.  In FIG \ref{FIG4} one can see that $\tilde{f}=(H/M)^\lambda$ satisfies $(H/M)^\lambda\gtrsim 1$, which is in agreement with equations \eqref{ns_rm} and \eqref{n_s_C}. Thus, the strong limit  $\tilde{f}=(H/M)^\lambda \gg 1$ is disfavored by observations.


\section{Concluding Remarks}\label{conclusion_f}

In the present paper, we study inflation and the generation of primordial fluctuations in the context of $f(R)$ Gravity’s Rainbow. After developing the general setup to study slow-roll inflation in these theories, we calculated the scalar and the tensor primordial power spectra generated during inflation. We assumed that the two energy-dependent rainbow functions $\tilde{f}$ and $\tilde{g}$ arising in the effective spacetime metric are time dependent. This is a natural assumption since the energy of the probe particles can depend on time for an expanding universe \cite{Ling:2006az}. 

Any viable inflationary model based on $f(R)$ gravity must be close to Starobinsky model \cite{Ketov:2017aau}. For instance,  one can write $f(R)=R+R^2 \mathcal{A}(R)$ where $\mathcal{A}(R)$ is a slowly varying function satisfying $\left|\mathcal{A}'(R)\right|\ll \mathcal{A}(R)/R$ and $\left|\mathcal{A}''(R)\right|\ll \mathcal{A}(R)/R^2$ \cite{DeFelice:2010aj}. For this class of models we have shown that gravity's rainbow can lead to new contributions on both inflationary observables $n_s$ and $r$. Although up to first order in slow-roll approximation there are not contributions coming from $\tilde{g}$, we found new imprints arising through the slow-roll parameter $\delta_{\tilde{f}}=\dot{\tilde{f}}/(H\tilde{f})$ associated with $\tilde{f}$. 

Since the rainbow functions can be time dependent, is totally reasonable to assume the ansatz $\tilde{f}(H)=(H/M)^\lambda$ where $\lambda$ is constant and $M$ is the mass scale of inflation \cite{Chatrabhuti:2015mws,Waeming:2020rir}. For instance, for $f(R)=R+(R/M)^2/6$, we obtained $n_s\simeq 1-2\left(1+\lambda\right)/{N_{k}}$ and $r\simeq 12 \left(1+\lambda\right)^2/N_{k}^2$. The standard results from Starobinsky model are recovered for $\lambda=0$ \cite{Starobinsky, DeFelice:2010aj}. Thus, using the latest observational constraints on $n_s$ and $r$ \cite{Planck:2018jri,BICEPKeck:2021gln} we found that small values of $\lambda$ ($\lambda<1$) are preferred. The observational bounds on $n_s$ put the strongest constraint on the values of $\lambda$ to give a nearly scale-invariant power spectrum. Higher values of $\lambda$ can also give results compatible with observations, but this would require higher values of the $e$-folding number of inflation $N_k$. Otherwise, higher values of $\lambda$ could violate the slow-roll conditions. Also, this could conflict with the expected results from the subsequent reheating era after inflation \cite{Kofman:1994rk,Kofman:1997yn,Bassett:2005xm,Lopez:2021agu}. 

Therefore, we conclude that the strong limit  $\tilde{f}=(H/M)^\lambda \gg 1$ is disfavored by observations. This result differs from what was obtained in Refs. \cite{Chatrabhuti:2015mws,Waeming:2020rir}. It is important to note that Eq. \eqref{H_t} satisfies $\ddot{H}=\mathcal{O}(\epsilon^2)$, and $\ddot{H}/(H \dot{H})=2 \lambda \epsilon$, with $\epsilon=-\dot{H}/(H^2)$. For Starobinsky inflation one has $\lambda=0$ and then $\ddot{H}/(H \dot{H})=0$.  But, in the presence of rainbow this is not zero and thus this term contributes to $n_s$, such as in Eq. \eqref{ns_rm}.


\begin{acknowledgments}
G. O acknowledges Dirección de Investigación, Postgrado y Transferencia Tecnológica de la
Universidad de Tarapacá for financial support through
Proyecto UTA Mayor 4723-22.
\end{acknowledgments}
\bibliographystyle{spphys}   
\bibliography{bio}   

\end{document}